\documentclass[12pt,preprint]{aastex}
\usepackage{amssymb,latexsym,graphics,eufrak}
\def\re{R_{e}}
\def\a0{a_0}

\def\vinf{V_{\infty}}

\begin{document}

\title{{\bf MOND and the ``Dearth of Dark Matter in
 Ordinary Elliptical Galaxies''}}
\author{Mordehai Milgrom$^1$ and Robert H. Sanders$^2$}
\affil{ $^1$Department of Condensed Matter Physics Weizmann
Institute\\
 $^2$Kapteyn Astronomical Institute, Groningen, NL}

\begin{abstract}
The recent findings of Romanowsky et al., of an ``unexpectedly''
small mass discrepancy within 5 effective radii in several
elliptical galaxies, are not surprising in the context of MOND. As
we show here, they are, in fact, in full concordance with its
predictions. One is dealing with high-surface-density galaxies
with mean accelerations rather larger than the acceleration
constant of MOND. These findings continue, and are now the extreme
examples of, the trend predicted by MOND:  the mass discrepancy
sets in at larger and larger scaled radii in galaxies with larger
and larger mean surface densities; or, equivalently, mean
accelerations.
\end{abstract}


 \section{Introduction}
 Romanowsky et al. (2003) have recently presented dynamical studies
of three elliptical galaxies based on the measurement of radial
velocities of a large number of planetary nebulae in each of these
galaxies. This method is rather more reliable, and is extended to
considerably larger radii, than the standard method of
interpreting the line-of-sight-integrated line profiles (see e.g.,
Gerhard et al. 2001,  and Baes and Dejonghe 2002). The dynamical
modelling of Romanowsky et al. (2003) ``indicates the presence of
little if any dark matter in these galaxies''. They conclude that
this does not naturally conform with the CDM paradigm.
\par
Our purpose is to point out that the results of Romanowsky et al.
are not only in agreement with the predictions of MOND, but go
beyond existing support of MOND, probing, as they do, the highest
acceleration end in the galaxy sequence, higher than what has been
probed with rotation curves of spiral galaxies.

MOND is an empirically motivated modification of Newtonian
dynamics at low accelerations, suggested as an alternative to dark
matter (Milgrom 1983). For quasi-spherical galaxies such as we
have at hand here, the true (MOND) acceleration $\bf g$ is, to a
very good approximation, related algebraically to the acceleration
$\bf g_n$ calculated from Newtonian dynamics via
$$ {\bf g}\mu(|{\bf g}|/\a0) = {\bf g_n},$$
where $\a0$ is the MOND acceleration constant, and the
interpolating function $\mu(x)$ approaches 1 in the limit $x\gg 1$
(the Newtonian limit), and approaches $x$ at low accelerations
 $x\ll 1$ (the MOND limit). Thus, in this low acceleration limit
$g=\sqrt{g_n \a0}$, predicting asymptotically flat rotation curves
for finite mass bodies, with an asymptotic velocity of $\vinf =
(GM\a0)^{1/4}$, where $M$ is the total (baryonic) mass.
\par
Newtonian dynamics predicts, of course, that masses having
homologous density distributions will have similar rotation
curves. This is not so in MOND, where the
existence of a preferred acceleration value implies different
behavior for galaxies with different surface densities relative to
the critical value $\Sigma_m = \a0/G$. In other words, the MOND
potential field, and, in particular, the shape of the MOND
rotation curve, depends on the parameter \cite{mil83}
$$\xi\equiv (MG/\re^2\a0)^{1/2}=\vinf^2/\re\a0, $$
where $\re$ is some measure of the size of the (baryonic) galaxy,
which for ellipticals we take as the effective radius (the
projected half-mass radius).

 Galaxy systems with $\xi\gg 1$ have internal accelerations
greater than $\a0$ in their main body, which is thus in the
Newtonian regime. Such high-surface-density objects are expected
to show only a mild mass discrepancy within a few $\re$, as, in
this case, $\xi\re$ marks the transition radius from the Newtonian
to the MOND regime. At the other end, low surface density
galaxies, with $\xi\ll 1$, are in the MOND regime throughout, and
are predicted to show a large mass discrepancy within the optical
image.

\par
As a class, normal elliptical galaxies lie at the high $\xi$ end
of this sequence.
 Assuming a mean $M/L=4$ for the elliptical galaxies observed and
catalogued by J{\o}rgensen et al. (1995a,b),
we estimate that $<\xi> = 3.5$ for this sample,
compared to values of typically 2 to 3 for HSB disk galaxies.
\par
In light of all this, the results of Romanowsky et al. (2003) are
expected in MOND.
\par
 In section 2 we compare these results with the predictions of
 MOND, and discuss the implications in section 3.

\section{Results}

\begin{figure}
 \plotone{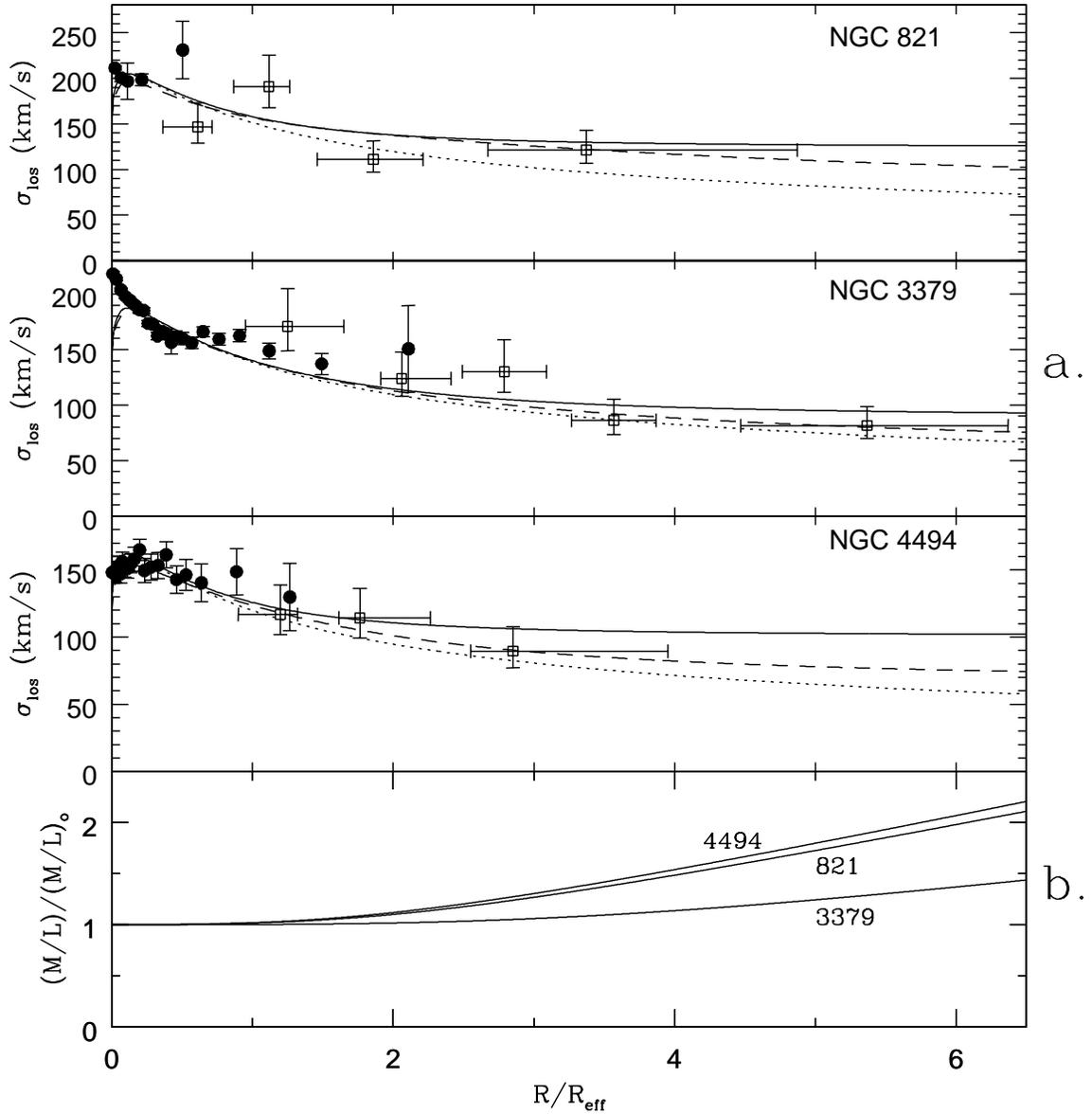}
 \caption{a.) The top three panels show the line-of-sight velocity
dispersion for the three elliptical galaxies observed by
Romanowsky et al. as a function of radius in units of the
effective radius.  The solid points are earlier observations of
the stellar velocity dispersion and the open squares are the PN
observations.  The dotted line is the predicted, Newtonian,
line-of-sight velocity dispersion for isotropic, constant $M/L$
models.  The solid lines are the MOND predictions for the same
models.  The dashed lines are the MOND predictions for models with
a variable anisotropy ratio, as described in the text. b.) The
lower panel shows the MOND prediction for the Newtonically deduced
runs of $M/L$ (normalized to the central value).}
\end{figure}

  The essential results of Romanowsky et al. (2003) are reproduced in
  our Fig.\ 1a, which is based on their Fig.\ 4.
   The points with error bars show
the velocity dispersion of planetary nebulae averaged over the
indicated radial bins as a function of radius in units of the
effective radius. The solid points are measurements of the stellar
velocity dispersion. We see that the planetary nebulae kinematics
probe the gravitational field to considerably larger radii than do
the stellar kinematics, particularly in the case of NGC 3379. The
dotted line, also from Fig. 4 of Romanowsky et al., shows the
predicted line-of-sight velocity dispersion assuming that light
traces mass and that the velocity distribution is isotropic. The
light distribution is fitted with a Hernquist  model (Hernquist
1990) having the appropriately chosen effective radius.  The
corresponding B-band mass-to-light ratios are, in solar units,
11.4 for NGC 821, 4.7 for NGC 3379, and 5.4 for NGC 4494.  The
obvious conclusion is that, except perhaps for NGC 821,  the
observations are consistent with these light-traces-mass models
and the required M/L values suggest little dark matter, if any. A
fourth galaxy, NGC 4697, was studied earlier \cite{mendez01}. It
was probed to smaller radii, and seems to indicate similar
behavior. We will only discuss it in passing.

\par
We also show in Fig.\ 1a the predictions of MOND, with the solid
lines corresponding to the same Hernquist models and to isotropic
velocity distributions. The MOND interpolating function is taken
to have the same form always used before in rotation curve
studies, i.e.,
$$\mu(x) = x/\sqrt{1+x^2},$$
and we use a value of $\a0=1\times 10^{-8}$ cm s$^{-2}$, based
upon the analysis of Bottema et al. (2002) making use of the new
distance scale. (All along we use the distances adopted by
Romanowsky et al.; recall that the distance affects the MOND
predictions differently from those of Newtonian dynamics.)
\par
 There is, of course, nothing preferable about an isotropic
velocity distribution; so, to gain some idea of the uncertainty
due the unknown orbit population,  we also show MOND models with a
variable anisotropy ratio (becoming more radial outwards). The
anisotropy parameter in these models is given by
$$\beta = {{r^2}\over{({r_a}^2 + r^2)}}$$
where the anisotropy radius $r_a = 3\re$ in all cases. Such a run
of anisotropy is typical of systems that form by dissipationless
collapse (van Albada 1982).
\par
Fig.\ 1b shows the MOND predicted $M/L$ runs, normalized to the
central value, for the three galaxies (calculated solely from the
light distribution, and simply given by $1/\mu(|{\bf g}|/\a0)$).
They give the predicted runs of Newtonically deduced mass
discrepancy. So, MOND does predict a mild mass discrepancy in
these three galaxy within the range studied. Romanowsky (private
communication) finds that the range of $M/L$ values deduced from
Jeans modelling of the velocity data of NGC 821 and NGC 4494, {\it
considering only models with a constant anisotropy
 ratio}, are milder than those given in Fig.\ 1b. The differences are,
however, well within the uncertainties due to measurement errors,
assumed distances, and modelling.
\begin{figure}
 \plotone{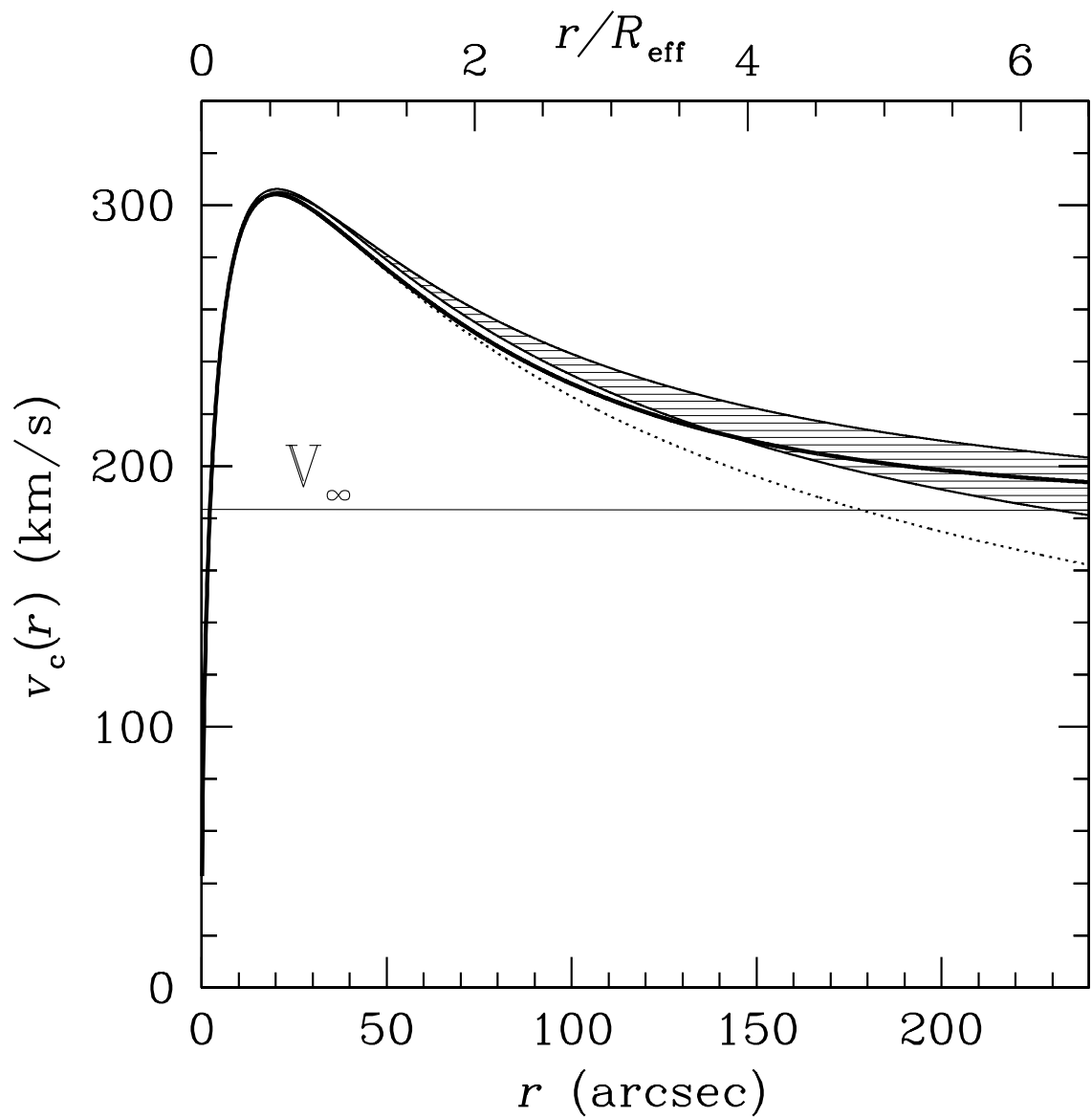}
 \caption{ The inferred rotation curves of NGC 3379.  The shaded area
shows the region permitted by the orbit modelling of PN radial
velocities by Romanowsky et al..  The dotted line shows the
constant $M/L = 5.5$, Newtonian rotation curve inferred from the
stellar light distribution.  The heavy solid line is the MOND
rotation curve determined from this constant $M/L$ model.  The
horizontal line shows the MOND asymptotic velocity ($\vinf$),
i.e., the circular velocity at infinity.  This could be compared
with the $\xi=5$ curve in Fig.\ 1b of Milgrom (1983)}
\end{figure}

\par
We see that, by and large, the binned, line-of-sight dispersion
data are as consistent with the MOND predictions as they are with
the Newtonian, no DM ones. The reason is that within the radii
measured the mass discrepancy predicted by MOND is not larger than
the measurement errors and/or the uncertainties due to the unknown
orbit population.
\par
To better manage these latter uncertainties, Romanowsky et al.
applied to the best studied galaxy, NGC 3379, an orbit-library
method described in detail in their paper (this entails, of
course, variable anisotropies in the velocity distribution). The
outcome of this analysis is a better constrained range of allowed
potential fields for this galaxy. This is described as a permitted
range of rotation curves given in their Fig\ A2, and reproduced
here in our Fig.\ 2 as the hatched area. The dotted curve in Fig.\
2 shows the Newtonian rotation curve for the adopted Hernquist
model, with a constant $M/L_B$ value of 5.5 solar units, chosen to
fit the inner parts, near the maximum (and no dark matter). The
solid line is the MOND prediction for the same mass model. The
MOND rotation curve for a spherical galaxy with a de Vaucouleur
profile having $\xi=5$ is, in fact, given in Figure 1b of Milgrom
(1983), and matches very nearly the deduced rotation curve for NGC
3379, which has $\xi\approx 5.7$.
\par
We see that this more refined distillation of the data, encapsuled
in the deduced rotation curve, does indicate the development of a
mild mass discrepancy at the larger radii, rather as predicted by
MOND. Note that because the accelerations at several effective
radii are near $\a0$, the exact shape of the MOND rotation curve
depends upon the precise form of $\mu$, but here this is a
relatively unimportant detail.  The main point is that MOND
predicts the very pronounced decline in the deduced rotation curve
of NGC 3379, as never seen before.

\section{Discussion}

We see that the MOND predictions of the kinematics of the galaxies
under study, as deduced from the observed light distribution
assuming a constant $M/L$ value, agree well with the observations,
assuming very reasonable orbit populations. The strength and
importance of these findings is, however, not in the exact quality
of the agreement. We have seen more impressive performances of
MOND in predicting rotation curves of disc galaxies (e.g., Sanders
\& McGaugh 2002), where the data are more accurate, and where the
uncertainty in orbit population is hardly present. The importance
here is that this new comparison confirms the prediction of MOND
regarding the high acceleration end of the galaxy distribution
\cite{mil83}, which has not been probed before.
\par
NGC 3379 is the best studied
of the three and with the measurements going the farthest in unit
of $\re$. It has $\xi\approx 5.7$, and seems to be the present
record holder among galaxies with measured extended rotation
curves. Accordingly, in conformity with the MOND predictions, it
is also the record holder in the extent of the observed decline of
its rotation curve.
 For NGC 821 $\xi\approx 3.6$, and for NGC 4494 $\xi\approx 3.4$.
For the previously studied NGC 4697, $\xi\approx 3.3$, similar to
that of NGC 4494.
\par
These last three galaxies are nearer in their $\xi$ value to some
of the HSB spirals, especially those with a substantial bulge. In
comparison, a similarly defined quantity for dwarf spheroidal
galaxies can be as small as $\xi\sim 0.1$. For LSB spirals
$\xi\sim 0.5$, (e.g., NGC 1560) and for HSB spirals with measured
extended rotation curves $\xi\sim 2.5$ (e.g., NGC 2903). In such
HSB spirals we expect to find a similar ``dearth of DM'' if we
measure their rotation curve only as far as a few half mass radii
(for NGC 2903, about 3.5 kpc). Such HSB galaxies are well fit by a
``maximum disk'', whereby the rotation curve in the inner regions
 is explained
by the visible disk with a reasonable $M/L$ ratio, implying that
very little dark matter is required there. Because the rotation
curves are, in fact, measured to rather larger radii, and a mass
discrepancy does develop in full strength at the outskirts, we
don't view these HSB galaxies as devoid of a mass discrepancy,
only as developing it at larger radii. This is also what MOND
predicts for the galaxies in the present study (as shown by the
$M/L$ runs in Fig. 1b).

\par
There are ellipticals with $\xi\sim 3$, as studied here, for which
a mass discrepancy has been claimed already at smaller radii. If
these are substantiated, and correctly interpreted, they would
certainly argue against MOND. There are, for example, ellipticals
such as M87 (e.g. C\^ot\'e et al. 2001) and M49 (e.g., C\^ot\'e et
al. 2003) that have been dynamically studied by both globular
cluster (GCs) dynamics and x-rays. They sit, however, at the
center of galaxy clusters (``Virgo A'' for M87, and the sub
cluster ``Virgo B'' for M49). It is known that, in the context of
MOND, there is a remaining ``missing mass'' problem for the
central regions of clusters (see, e.g., Sanders 2003); that
additional mass, perhaps undetected baryonic matter or massive
neutrinos, is required to make up the MOND dynamical mass budget.
The mass discrepancies deduced around such galaxies may be
reflecting these known cluster discrepancies. Moreover, in the
case of M49, which presumably falls in the external field of the
main cluster together with ``Virgo B'',  there is a possibility of
departure from viriality in the globular cluster system (magnified
by MOND effects). This might be suggested by two peculiarities: a.
the velocity dispersion of the metal-rich GCs is significantly
lower than that of the metal-poor ones although they both
supposedly probe the same gravitation field. In fact, if taken in
itself, the metal-rich-GC dispersion is consistent with MOND. b.
the GC velocity dispersion profile adopted by C\^ot\'e et al.
(2003) (taken as that of the combined population) is not a
continuation of the stellar dispersion profile at smaller radii,
which, like that of the metal-rich GCs, does appear to decline
with radius, and does not imply a mass discrepancy. In any event,
the examples of M87 and M49 are not the clean cases of isolated
galaxies which would comprise ideal tests of MOND.

\par
In another instance, Gerhard et al. (2001) studied the dynamics of
 ellipticals by modelling the profiles of the
line-of-sight-integrated lines, extending, typically, to
$0.5\re-2\re$. Their deduced, enclosed $M/L$ values rise slowly
with radius. The increase in $M/L$ at the last measured point over
the central value is in most cases between 10 and 50 percent, with
a few going up to 80 percent. Gerhard et al. take this rise to be
the onset of the mass discrepancy in these galaxies, which would
thus occur at rather smaller radii than found by Romanowsky et al.
(2003). Gerhard et al. do not show errors on their deduce values,
but they state that ``the typical uncertainty in the outermost
circular velocity is $\pm(10-15)\%$''. This translates to a
$\pm(20-30)\%$ uncertainty in $M/L$, rather comparable with the
whole claimed effect in most cases. When plotted against the
acceleration, their $M/L$ values begin to increase at
accelerations between 3 and 30 times higher than $\a0$. Gerhard et
al. view this as showing that the mass discrepancy in ellipticals
sets in at accelerations that are an order of magnitude higher
than in spirals, which, if true, would fly in the face of MOND.

\par
We contend that these M/L increases reported by Gerhard et al.,
whatever they are due to, do not mark the onset of the mass
discrepancy. Two of the three galaxies studied in Romanowsky et
al., NGC 4494 and NGC 3379, are also in the Gerhard et al. sample.
For NGC 4494, Gerhard et al. find that $M/L$ increases already by
40 percent at their last measured point of $0.7\re$. For NGC 3379,
they deduce a rotation curve that becomes flat at about $1\re$ and
remains so to their last measured point of about $2\re$, where
their deduced $M/L$ value has risen already to 20 percent above
the central value, whereas Romanowsky et al. tell us that the
rotation curve declines at least down to $6\re$. If the results of
Romanowsky et al. are valid, the $M/L$ increases that Gerhard et
al. find for these two galaxies cannot possibly mark the onset of
the mass discrepancy; and this obviously casts heavy doubt on
their conclusion for other galaxies. Note also that their $M/L$
values increase much more slowly as a function of the acceleration
than they do in spirals: instead of increasing as $1/a$, as
predicted by MOND, and as observed, they increase by a few tens of
percents over a decade in $a$. So, this is certainly not MOND
translated to higher accelerations, or a ``non-universality of
$\a0$'' as has been claimed.

These deduced $M/L$ increases may result from actual increase in
the stellar $M/L$ values; or, they may be artifacts of the
analysis, or due to unaccounted for systematics, as, in fact, has
been proposed by Baes and Dejonghe (2002).

 \acknowledgements
We thank Aaron Romanowsky and Nigel Douglas for providing their
data in practical form, and for very helpful discussions.

\clearpage
\end{document}